\documentclass[twocolumn,floatfix,prl,aps,showpacs,superscriptaddress]{revtex4-1}
\usepackage{epsfig,graphicx,tabularx,color,amsmath}

\begin{document}
%
\title{Phase-slips and vortex dynamics \\in  Josephson oscillations between Bose-Einstein condensates}

\author{M. Abad}
\affiliation{INO-CNR BEC Center and Dipartimento di Fisica, Universit\`a di Trento, 38123 Povo, Italy}
\author{M. Guilleumas}
\affiliation{Departament d'Estructura i Constituents de la Mat\`{e}ria  and IN2UB,\\
Facultat de F\'{\i}sica, Universitat de Barcelona, E--08028 Barcelona, Spain}
\author{R. Mayol}
\affiliation{Departament d'Estructura i Constituents de la Mat\`{e}ria and IN2UB,\\
Facultat de F\'{\i}sica, Universitat de Barcelona, E--08028 Barcelona, Spain}
\author{F. Piazza}
\affiliation{Technische Universit\"at M\"unchen, James-Franck-Stra{\ss}e 1, 85748 Garching, Germany}
\author{D. M. Jezek}
\affiliation{IFIBA-CONICET, Pabell\'on 1, Ciudad Universitaria, 1428, Buenos Aires, Argentina}
\author{A. Smerzi}
\affiliation{QSTAR, INO-CNR and LENS, Largo Enrico Fermi 2, 50125, Firenze, Italy}

\date{\today}
\begin{abstract}
We study the relation between Josephson dynamics and topological excitations in a dilute Bose-Einstein condensate confined in a double-well trap.
We show that the phase slips responsible for the self-trapping regime are created by vortex rings 
entering and annihilating inside the weak-link region or created at the center of the barrier and 
expanding outside the system. Large amplitude oscillations just before the onset of self-trapping are 
also 
strictly connected with the dynamics of vortex rings at the edges of the inter-well barrier. 
Our results extend and analyze the dynamics of the vortex-induced phase slippages suggested
 a few decades ago in relation to the ``ac" Josephson effect of superconducting and superfluid helium 
systems.
\end{abstract}
\maketitle

{\it Introduction.}
Phase coherence and superfluidity are two characteristic and most intriguing features of quantum fluids. Both phenomena are marked  by a
complex-valued order parameter. The nature of the order parameter is different in superconductors, helium superfluids or ultracold gases, but the most important common feature is a rigid (due to inter-particle interactions) local phase (arising from macroscopic coherence) whose gradient is proportional to the superfluid velocity. This describes a class of universal phenomena such as the quantization of vortices or the stability of persistent currents.

One of the most important and counterintuitive consequences of a non-vanishing order parameter with a non-linear dynamics is the Josephson effect between two superfluid bulks.
In particular, Bosonic Josephson Junctions (BJJs) created with two weakly linked, dilute Bose-Einstein 
condensates,
are characterized by a rich class of dynamical regimes \cite{smerzi97}. 
In the ``plasma" oscillations, both the relative population and the relative phase
between the two bulks oscillate sinusoidally at a frequency decreasing with the height of the tunneling barrier renormalized by the strength of the interatomic interaction.
In the ``ac" Josephson regime, the relative phase increases in time due to an external
force (induced for instance by a linear potential superimposed to a symmetric double-well). In the ``macroscopic quantum self-trapping" regime (MQST), the relative phase increases in time driven by a difference in internal energy between the two bulk regions of the weakly linked system.
The internal energy difference is induced by a self-sustained relative population imbalance due to the intrinsic non-linearity of the system. Moreover, $\pi$ modes 
with a relative phase oscillating around an average value $\pi$ are also possible. 
Most of these effects have been recently experimentally observed
with Bose-Einstein condensates (BECs) trapped in double-well~\cite{albiez05,levy07,leblanc11} and optical lattice potentials~\cite{Cataliotti2001}, in the internal BEC dynamics between two hyperfine levels~\cite{Zibold2010} as well as in different physical systems like two weakly coupled polariton condensates \cite{polariton}.

The ``ac" and the ``MQST" regimes are both characterized by a ``running" relative phase monotonically increasing with time. 
The phase is of course single valued. In particular, when it is calculated in a closed loop around a zero-density
region of the order parameter the phase can return to its original value $\pm 2 n \pi$.
This leads to the concept of phase slippage induced by the creation of a singly quantized vortex crossing the barrier or the bridge weakly linking two superfluid bulks, introduced to explain the ``ac" Josephson effects of superfluid Helium \cite{Anderson1966}.
An experimental signature of phase-slips during Josephson oscillations has been found as jumps of the macroscopic current between two weakly linked 
superfluid Helium reservoirs driven by an external induced pressure difference \cite{Hoskinson2006}. However, this behavior was observed only far outside the Josephson tunneling regime, whereas in the latter the current shows smooth sinusoidal oscillations.


In this work, we study the dynamics of phase-slippage in the macroscopic quantum self-trapping regime of two weakly linked Bose-Einstein condensates.
We focus on a 3D case and show that the topological excitation responsible for the phase slippage is a 
vortex ring crossing the junction by radially shrinking (expanding) to (from) a point at the center. 
We also observe a non-trivial vortex dynamics just before the onset of MQST, where the large amplitude 
population oscillations are driven by vortices floating in the low density barrier region.
The creation of topological excitations and their dynamics are both very fast compared 
to the relative phase and population dynamics among the two bulk regions. This justifies the use of a two-mode approximation
(TMA), where the dynamics is described only in terms of collective variables.
In contrast to the full numerical solution, in the TMA the phase slippage takes place through the creation of a dark soliton-like structure which creates a zero-density surface across the barrier. TMA also, somehow surprisingly, reproduces well the sudden jump in the value of the local superfluid velocity inside the barrier induced by a phase-slip. 
Taking into account a correction factor that renormalizes the interaction term~\cite{Jezek2013}, we show that 
the predictions of the TMA are also quantitative.

Phase slips are also a mechanism for superflow decay in the ``dc'' Josephson experiment. In the context of BEC, they have been predicted~\cite{piazza2009,piazza2011} to occur when the condensate flows through a barrier at a velocity larger than a critical value, and observed experimentally as jumps in the macroscopic current~\cite{Ramanathan2011,hadz_2012,Wright2013a,Wright2013b,jend2014} . The creation of vortices and their relation to self-trapping dynamics in a dipolar self-induced Josephson junction was discussed in \cite{abad11b}.

\begin{figure}[t]
\epsfig{file=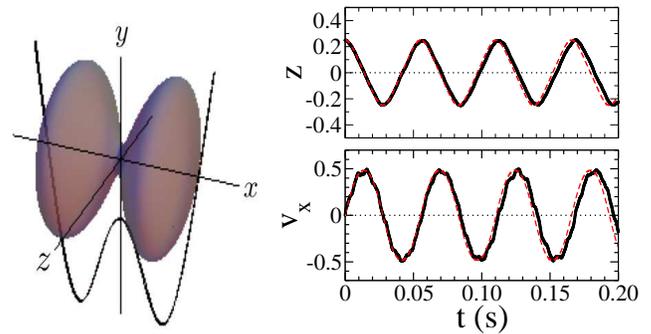, height=0.48\linewidth,width=0.42\linewidth, clip=true}\hspace{1ex}
\epsfig{file=fig1b.eps, width=0.55\linewidth, clip=true}
\caption{Left: Scheme of the double well potential and the geometry of the system. Right: Time evolution of the population imbalance (top) and local velocity at the junction (bottom, arbitrary units) for 
$Z_0=0.25$. Solid lines are numerical GPE results, and dashed lines TMA.}
\label{figjos}
\end{figure}

{\it The model}.
Compared to superfluid helium, the advantage of studying the phase slippage mechanism in BECs is that 
their microscopic dynamics is quantitatively well described by 
the time-dependent Gross-Pitaevskii equation (GPE) 
\begin{equation}
i\hbar\frac{\partial\psi({\mathbf r},t)}{\partial t}=
\left[-\frac{ \hbar^2 }{2 m}{\bf \nabla}^2  +
V({\mathbf r})+g|\psi({\mathbf r},t)|^2\right]\psi({\mathbf r},t) \,,
\label{tdgp}
\end{equation}
where $m$ is the atomic mass, $V$ is the double-well external potential shown in Fig.~\ref{figjos}, $g$ is the interaction coupling constant giving rise to nonlinearity, and $\psi$ is the wave function or order parameter. 
The local dynamics will be mostly contained in the phase, $\phi(\mathbf{r})$, of the order parameter, which in terms of hydrodynamic variables is written as $\psi=\sqrt{n}e^{i\phi}$, with $n(\mathbf{r})$ the density. A relevant quantity describing the microscopic dynamics is
the velocity field of the BEC projected along the junction axis (which we label here as $x$ axis) and integrated around the junction region:
\begin{equation}
 v_x=\frac{1}{A_{junc}}\int_\text{junc} \mathbf{v}(x,y,0)\cdot \hat{x}\,dx \,dy  \,,\label{eqvelo}
\end{equation}
where the local superfluid velocity $\mathbf{v}=\hbar/m \nabla \phi$ and the integral is computed at the barrier region centered at $x=y=0$. We take the plane $z=0$ for convenience, but a 3D integral would give an equivalent result. The quantity $A_{junc}$ is the area of the junction over which the integral is evaluated. As we shall see, $v_x$ is sensitive to the presence of topological excitations.
The relevant macroscopic quantity is instead the population imbalance $Z$ between the regions lying on the two sides of the junction:
\begin{equation}
Z=\frac{1}{N}\left[\int_{-\infty}^0\!\!\!\!\!\!\!\!dx\int_{-\infty}^\infty\!\!\!\!\!\!\!\!dy\,dz|\psi|^2-\int_{0}^\infty\!\!\!\!\!\!\!\!dx\int_{-\infty}^\infty\!\!\!\!\!\!\!\!dy\,dz|\psi|^2\right]\;,
\end{equation}
where $N$ is the total atom number.
The external potential is similar to the one used in the experiments~\cite{albiez05}, $ V(x,y,z) = (1/2) m \omega^2 r^2 + V_0 \cos^2(\pi x/q_0)$, with $V_0$ the strength of the optical lattice, $q_0$ its wavevector, and $r^2=x^2+y^2+z^2$. 
The interatomic interactions are characterized by the scattering length $a$, related to the coupling constant as $g=4 \pi \hbar^2 a/m$. 
We initialize the system with the ground state wave function of a tilted double well (see \cite{abad11} for more details), which gives initial conditions $Z_0$ and $\phi_0(\mathbf{r})=0$ everywhere. At $t=0$ the tilting potential is switched off and Eq.~(\ref{tdgp}) is solved. For concreteness, we have considered a $^{87}$Rb condensate with parameters: $a=100.87\, a_B$, $ \omega = 2 \pi~ \times 70$~Hz, $V_0/\hbar=2\pi\times 413$~Hz  and $q_0 = 5.2 \, \mu$m. The total number of atoms is $N=1150$. In this configuration the critical imbalance that gives the onset of self-trapping is $Z_c\simeq 0.33$. 

{\it Local dynamics close to the pendulum instability}. 
In the two-mode model, the population imbalance during small amplitude (or ``plasma") oscillations 
evolves sinusoidally in time.
The right panels of Fig.~\ref{figjos} show the GPE numerical evolution of population and velocity at the junction with initial condition $Z_0=0.25$ (solid lines).  
The imbalance evolves almost sinusoidally in time and the local velocity is shifted by $\pi/2$.

\begin{figure}[t]
\epsfig{file=fig2a.eps, width=0.52\linewidth, clip=true}%
\epsfig{file=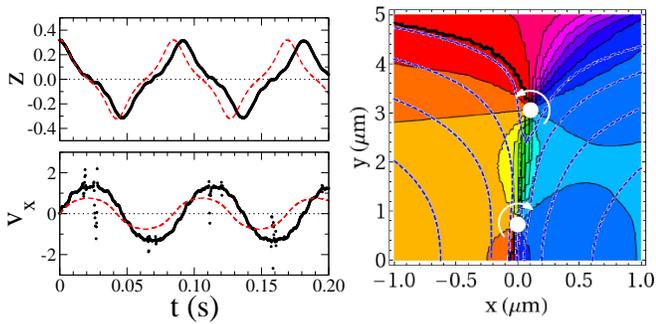, width=0.48\linewidth, clip=true}%
\caption{Left: Time evolution of the population imbalance (top) and local velocity (bottom, arbitrary 
units) for $Z_0=0.32$. Solid lines (dots) are numerical GPE results, and dashed lines TMA. Right: Cut of 
the phase at $z=0$ for $t=111.80$~ms. The dashed lines correspond to the equidensity 
contour at $10^{-k} n_{\rm max}$, with $k=4,3,2,1$ (inwards) and $n_{\rm max}$ being the maximum density 
in the plane $z=0$. As a guide to the eye, the projection of the vortex rings on the plane $z=0$ is 
marked with a circle and the corresponding circulation with an arrow. 
}
\label{figcrit}
\end{figure}

By increasing the initial amplitude the oscillations become anharmonic:
this is simply captured by the pendulum analogy of the BJJ dynamics \cite{smerzi97}.
When the phase approaches its maximum value $\pi$ (the pendulum getting close to the upright position) and the atomic flow is reversed,
rugosities appear in the zero crossings of the population imbalance
and the local velocity shows very sharp peaks 
as can be seen in the left panels of Fig.~\ref{figcrit}. 

These peaks are caused by
vortex rings entering the junction region but not crossing it completely, thus not giving rise to a $2\pi$ phase-slip. 
Since our configuration shows axial symmety around the $x$ axis, these topological excitations are  circular vortex rings, which are vorticity lines along a circumference lying in the $x=0$ plane, appearing as a vortex-antivortex pair in the $z=0$ plane. In the right panel of Fig.~\ref{figcrit} the local phase $\phi(\mathbf{r})$ (for $z=0$ and $y>0$) shows singularities in correspondence to the vortex rings.
\begin{figure}[t]
\epsfig{file=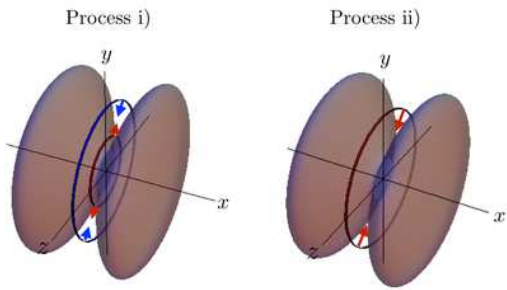, width=0.82\linewidth}
\caption{Sketch of the two possible processes characterizing the vortex ring dynamics.}
\label{fig_dyn}
\end{figure}
As illustrated in Fig.~\ref{fig_dyn}, there are two different processes that can take place: (i) a 
vortex ring is created at the center and tries to cross the junction by expanding and (ii) a vortex ring 
is created at the surface and tries to penetrate the barrier by shrinking. We have seen both processes 
in the simulations, but only the local phase for process (i) is shown in the right panel of 
Fig.~\ref{figcrit}, corresponding to the second velocity maximum. 
The lower phase singularity belongs to the inner vortex ring expanding outward while the upper singularity belongs to a second vortex ring with opposite vorticity external to the junction (seen as an anti-vortex in the $z=0$ plane). As will be discussed later for the MQST regime, these secondary vortices have the role of helping the expanding vortices to fully cross the junction.


{\it Local dynamics in self-trapping regime}. 
\begin{figure*}
\epsfig{file=fig4a.eps, width=0.3\linewidth, clip=true}%
\epsfig{file=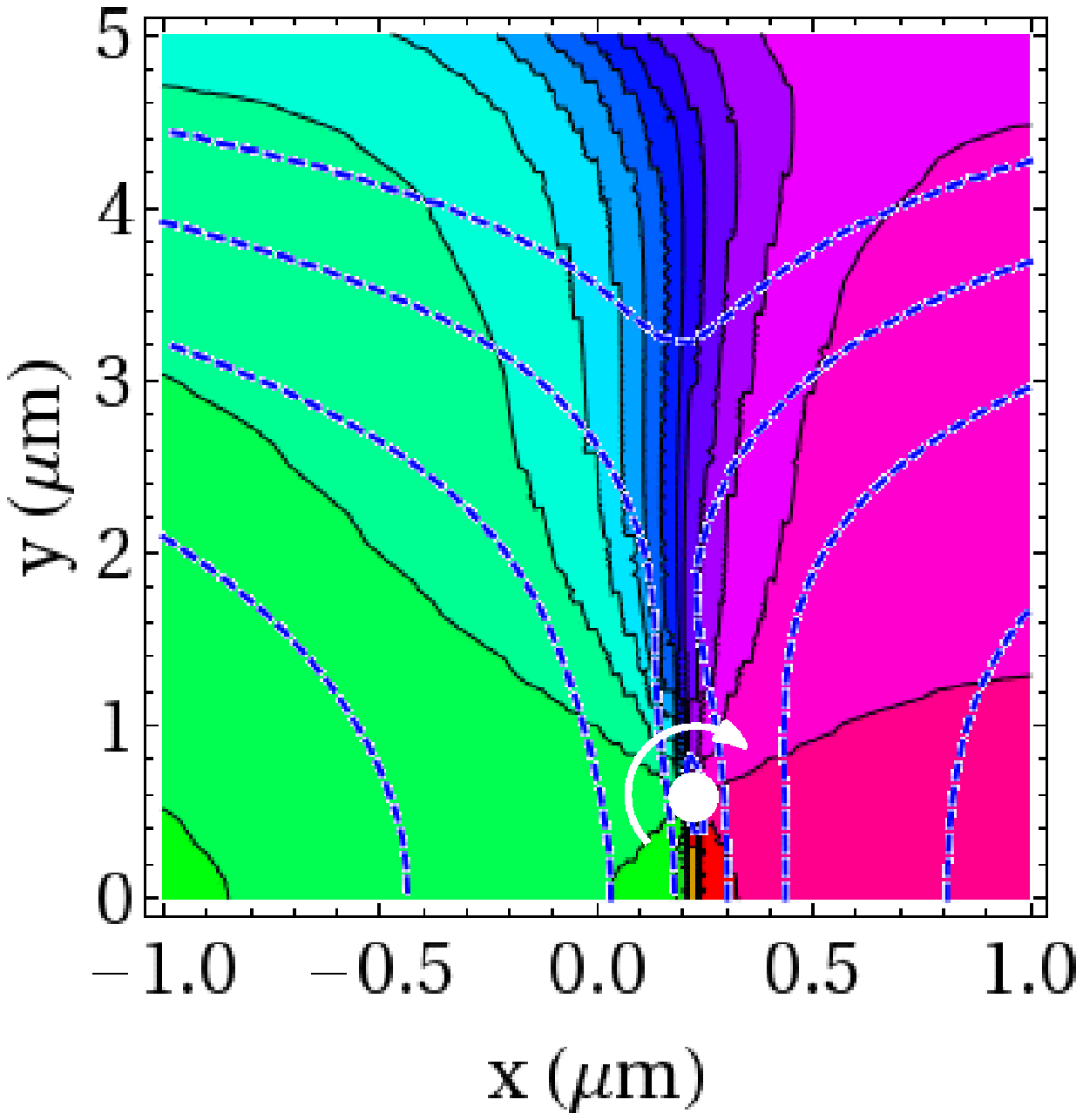, width=0.23\linewidth, clip=true}%
\epsfig{file=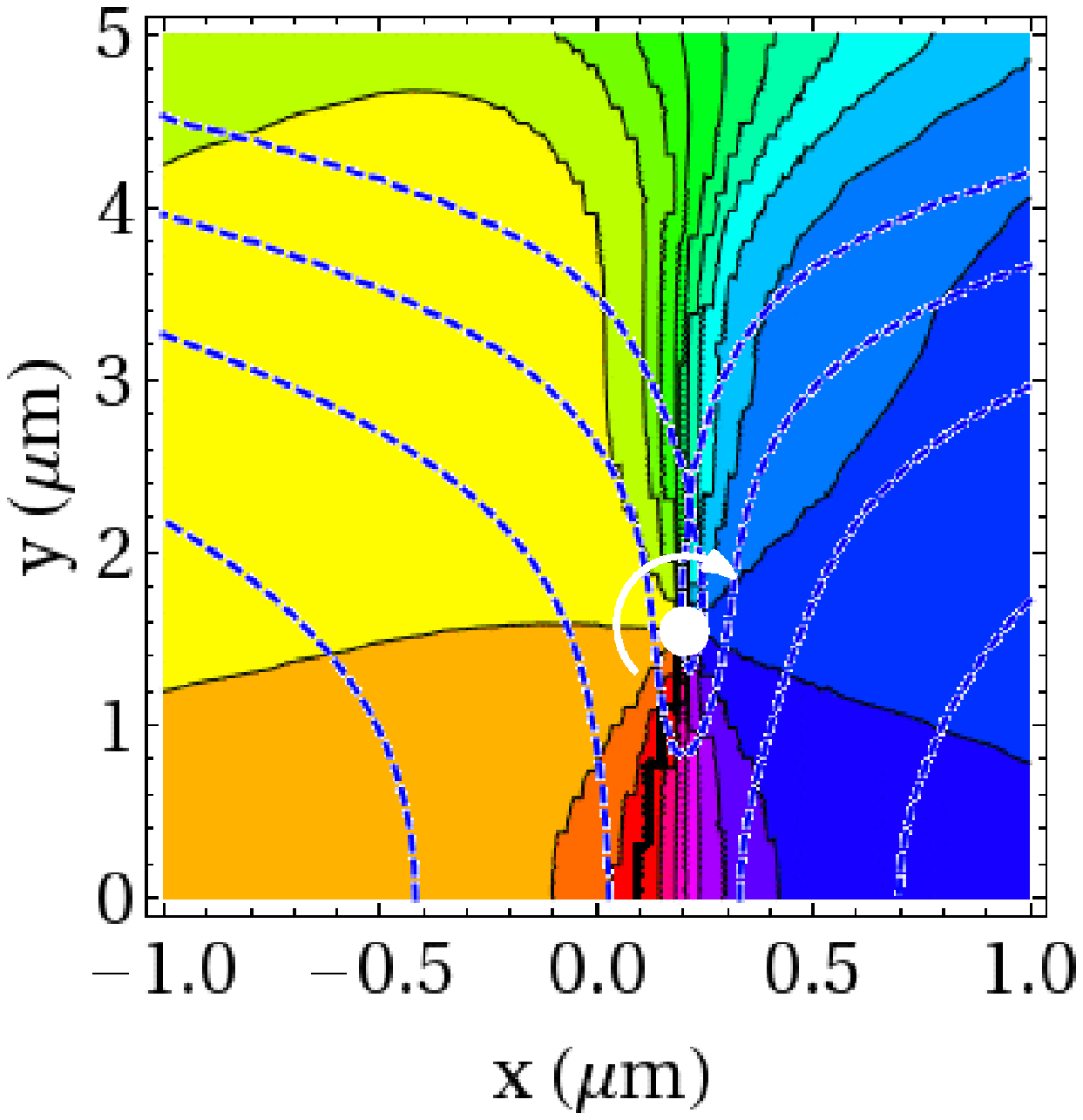, width=0.23\linewidth, clip=true}%
\epsfig{file=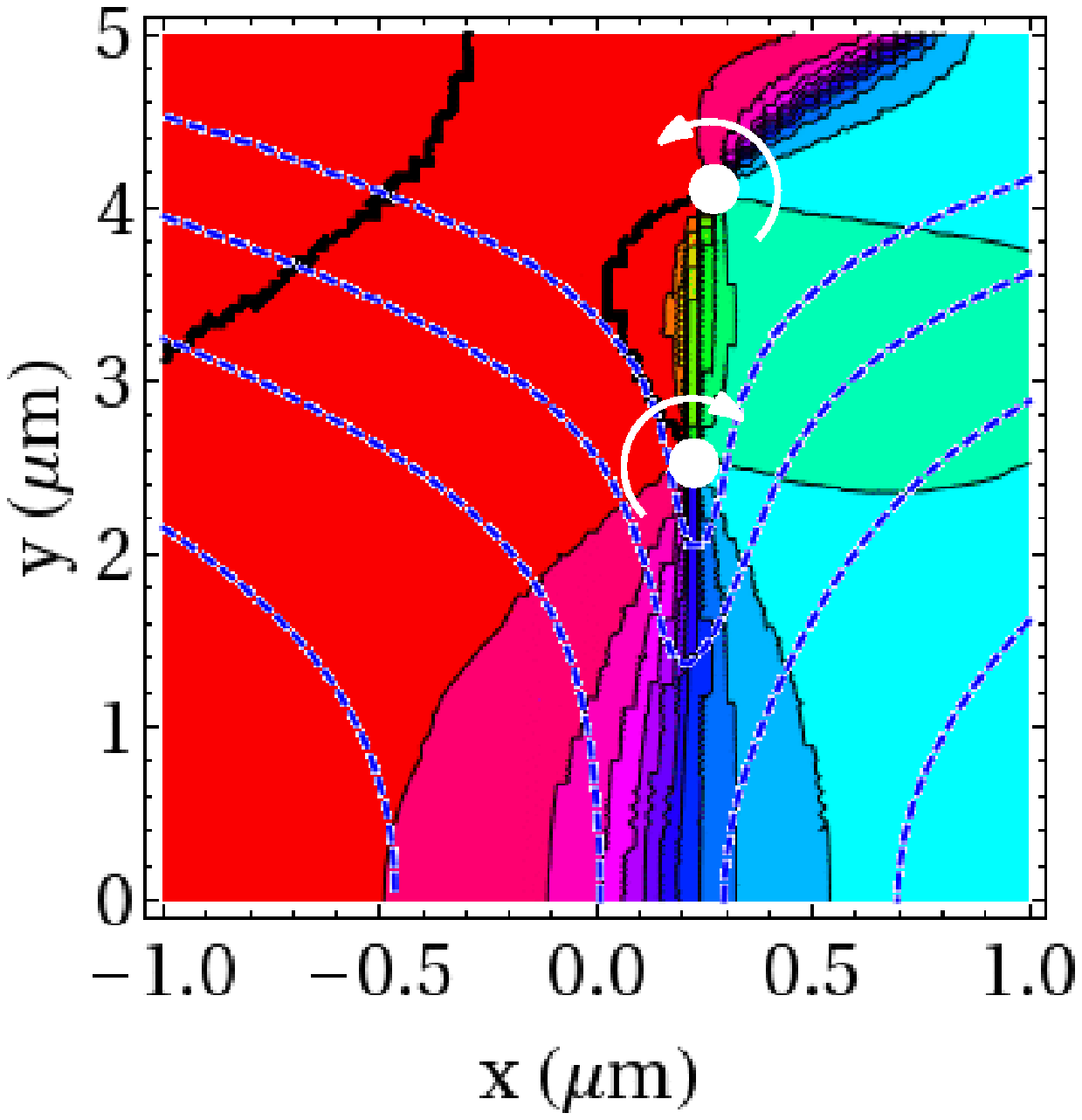, width=0.23\linewidth, clip=true}%
\caption{ Time evolution of the population imbalance (top left) and velocity at the junction (bottom, 
arbitrary units) for $Z_0=0.34$. Solid lines (dots) are numerical GPE results, and dashed lines TMA. 
Phase snapshots at the $z=0$ plane around the first imbalance minimum. The times from left to right 
correspond to $ t=16.6$ ms, $17.2$ ms, and $17.6$ ms.  }
\label{figst}
\end{figure*}
Above the critical value of the initial imbalance
the average imbalance remains locked at positive values and the phase grows linearly in time. Taking again the pendulum analogy, this corresponds to a situation where the pendulum no longer oscillates but it keeps going about in circles with the pendulum angle (the relative phase) steadily increasing with time. 
In this case, the velocity is maximum at the equilibrium position (minimum population imbalance). 

The linear growth of the phase difference can be understood in terms of phase slips, which are reflected 
in sharp decreases of the local velocity when the imbalance has a minimum, as seen in the left panels of 
Fig.~\ref{figst}. These sharp decreases are caused by vortex rings that have now enough energy to 
completely cross the junction, leaving the $2\pi$ phase-slips behind. Also in this case both processes 
(i) and (ii) mentioned above are possible.  An example of vortex passage is shown in Fig.~\ref{figst}, 
for initial condition $Z_0=0.34$, where the three phase snapshots show a vortex (which results 
from the projection of the vorticity line of the $y>0$ part of the ring vortex on the plane $z=0$) 
crossing the junction from inside to outside. Notice that the junction does not lie exactly in the $x=0$ 
plane, but it is slightly shifted. This is an effect of the imbalance combined with the nonlinear 
interaction.

When the vortex ring is created within the barrier (process (i) above) it has to expand outwards to produce the phase slip. In this process the ring increases its energy. In order to keep the total energy constant, a second ring with opposite circulation, an anti-vortex in the $z=0$ plane (seen in the rightmost panel of Fig.~\ref{figst} and the right panel of Fig.~\ref{figcrit}), comes in from the system's surface and annihilates the inner vortex. The annihilation takes place in the low density region outside the junction.

{\it Topological excitations and flow inversion}. 
The link between local flow inversion and topological excitations can be understood within a TMA 
\cite{smerzi97}. 
This model assumes that the condensate wave-function can be written as a superposition of the left
($\Phi_L(\mathbf{r})$) and right ($\Phi_R(\mathbf{r})$) real mode functions, localized in the corresponding wells, namely
\begin{equation}
 \Psi(\mathbf{r},t) = \sqrt{N_L(t)}e^{i\phi_L(t)}\Phi_L(\mathbf{r}) + \sqrt{N_R(t)}e^{i\phi_R(t)}\Phi_R(\mathbf{r}) \,,
\end{equation}
with $N_i$, $\phi_i$ the number of particles and phase in each of the wells, $i=L,R$. 
For a symmetric double well, the main features of the BJJ are described by the Hamiltonian 
$ H(Z,\Delta\phi)=\frac{1}{2}\Lambda Z^2 - \sqrt{1-Z^2}\cos\Delta\phi$~\cite{foot}, where the imbalance 
and the phase difference are defined, respectively, as $Z=(N_L-N_R)/N$ and  $\Delta\phi=\phi_R-\phi_L$. 
The parameter $\Lambda$ is proportional to the ratio between interactions and linear coupling, and it is the only parameter in this model. Depending on the initial conditions the Hamiltonian predicts different possible trajectories in the $Z-\Delta\phi$ phase space, which can be closed (like plasma oscillations and $\pi$-modes) or open (like MQST). 
The transition between these takes place at a critical value of the initial imbalance (for $\Delta\phi=0$) $Z_{c}$, which corresponds to the separatrix line between the closed and open orbits. 

In the TMA, the local velocity of the superfluid across the junction can be written as
\begin{equation}
 \mathbf{v}(\mathbf{r},t)=\frac{\hbar}{2m}\frac{N}{n(\mathbf{r},t)}\sqrt{1-Z^2}\sin(\Delta\phi)\left[ \Phi_L\nabla\Phi_R- \Phi_R\nabla\Phi_L \right]\ .
\end{equation}
This velocity is related to the total macroscopic current
\begin{equation}
 J=\int\mathbf{j}\cdot d\mathbf{S} \propto\sqrt{1-Z^2}\sin(\Delta\phi) \,, \label{EqJ}
\end{equation}
where we have used the hydrodynamic relation $\mathbf{j}(\mathbf{r},t)=n(\mathbf{r},t)\mathbf{v}(\mathbf{r},t)$. Equation~(\ref{EqJ}) shows that the junction is characterized by a sinusoidal current-phase relationship in all of its dynamical regimes. 

To characterize what happens inside the junction during flow inversion and to compare with GPE results, we can calculate the mean velocity at the junction, Eq.~(\ref{eqvelo}).
To obtain a very good quantitative agreement with the GPE simulations, we fit the value of $\Lambda$ according to \cite{Jezek2013} (see Supplemental Material).
The time evolution predicted by the TMA is shown as dashed lines in Figs.~\ref{figjos}, \ref{figcrit} 
and \ref{figst} together with the evolution of the imbalance, and compared with the numerical GPE 
results.
In particular, notice that in the MQST regime the TMA predicts, as in the GPE  analysis,
a sudden change in the sign of the local velocity. However, in the TMA the sharp decrease is not caused by
a vortex ring but by a (dark) soliton-like structure occurring when the relative bulk phase reaches $\Delta\phi = \pi$, since the modes $\Phi_L$ and $\Phi_R$ are static and do not evolve in time.

The linear increase in the local velocity with a sudden switch of sign at $\Delta\phi = \pi$ requires a 
precise interplay between the dynamics of the current and the density. In particular, while both decrease 
when approaching the critical point, the density decreases at a faster speed so as to provide a linear 
increase
of the local superfluid velocity $v_{s} = J / n$. This process continues up to the point where the density vanishes and the phase slip takes place. It is somehow surprising that such a delicate interplay can be captured by the two-mode approximation, where the spatial wave-functions have a frozen profile and the dynamics
is expressed only in terms of collective variables.

{\it Discussion and Conclusions}.
We have studied the relation between phase slips and topological excitations in a BJJ undergoing Josephson oscillations. In our 3D configuration, the dynamical regimes of the junction can be related to different dynamics of vortex rings. 
We expect that the nature of the topological excitations depends on the effective dimensionality of the system, being a soliton in 1D and a couple vortex-antivortex in 2D.
This zoology exists because, depending on the dimensionality, it can be energetically favorable to create a phase slip with an instantaneous cut between the two bulks through a dark soliton rather than with a shrinking  ring vortex. 
The net effect, in all cases, is the $2\pi$ phase slip. 
In MQST regime of the junction, the speed of the vortex crossing the barrier is quite higher than the
collective population/phase dynamics and can be considered instantaneous in the tunneling time scale. Therefore, as far as the collective variable dynamics is concerned,
the details of the topological excitations are not important and the TMA can provide a good approximate description of the system also in higher
dimensions. 

Compared to superfluid Helium, the advantage of studying the phase slippage mechanism in BECs is that 
their microscopic dynamics is quantitatively well described by the 
the time-dependent Gross-Pitaevskii equation. This dynamical analysis has allowed us to clarify that 
the phase slippage is the net effect of the annihilation of a 
(ring)vortex and an anti(ring)vortex or of a shrinking ring-vortex, and not a single topological excitation crossing the barrier region as previously suggested 
in the literature. Moreover, a BEC junction might allow the experimental observation of such topological excitations in the Josephson regime. Our results can also
clarify the role of topological excitations in the experimental observation of MQST in polariton systems \cite{polariton}.
To conclude, we notice that we have focused our analysis at the onset of the self-trapping regime but 
it will be interesting to investigate the phase slip dynamics of the ``ac'' regime as well as of the $\pi$-oscillations.

\acknowledgments
M. A. acknowledges financial support by ERC through the QGBE grant and by Provincia Autonoma di Trento.
M. G. and R. M. acknowledge financial support under Grant No. FIS2011-28617-C02-01 from MINECO (Spain),
Grant No. 2009SGR1289 from Generalitat de Catalunya (Spain). D. M. J. acknowledges financial support under Grant PIP 11420090100243 from CONICET (Argentina).
A. S. acknowledge financial support by the European Commission small or medium-scale focused research project QIBEC, Contract Nr. 284584.
QSTAR is the MPQ, LENS, IIT, UniFi Joint Center for Quantum Science and Technology in Arcetri.

\clearpage
\section*{Correction to the two-mode model}

In the self-trapping regime, the frequency of the two-mode approximation (TMA) 
does not agree perfectly well with full simulations of the Gross-Pitaevskii equation (GPE)
since the densities in
the left and right wells are different from balanced wavefunctions used to calculate the parameter $\Lambda$ \cite{Ananikian}.
Possible ways to obtain a good quantitative agreement are to make use of an effective
  interaction parameter as in 
\cite{Jezek2013} or to explicitly
consider the dependence of the local chemical potential on the number of particles in each well 
\cite{smerzi2003}.

Here we follow \cite{Jezek2013}, where the effect of the imbalance is taken into 
account in the two-mode 
model by renormalizing the interaction parameter $U$.
 To find the correction for a finite imbalance
$Z$ we first calculate the quantity (see \cite{Jezek2013} for details):
\begin{equation}
 \frac{U_R}{U}= \frac{1}{\int d\mathbf{r}\,(n(N_0))^2}\int d\mathbf{r}\, n(N_0) n(N_0+\Delta N)
\end{equation}
where $n(N_0)=|\Phi(N_0)|^2$ and $n(N_0+\Delta N)=|\Phi(N_0+\Delta N)|^2$ are, respectively,
the densities of the ground state of the double well with $N_0=1150$ and $N_0+\Delta N$ particles. 
\begin{figure}[b]\centering
 \epsfig{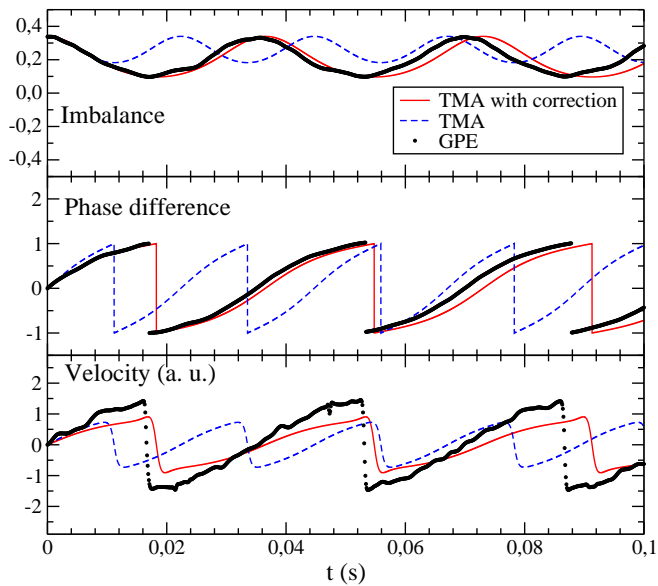}
 \caption{Comparison of two-mode model results (imbalance, phase difference and velocity at the junction) with GPE.
}\label{figmtm}
\end{figure}
Notice that both $\Phi(N_0)$ and $\Phi(N_0+\Delta N)$ are normalized to unity.
In this approximation, the new, effective interaction parameter $U_{eff}=(1-\alpha)U$ can be
 found using the  $\alpha$ value obtained  from the linear fit 
\begin{equation}
 \frac{U_R}{U}\simeq 1-\alpha\frac{\Delta N}{N_0} .
\end{equation}
For our system, using $U_{eff}$ leads a renormalized
two-mode model parameter $\Lambda=36.5$. 
Including this correction into the TMA (solid lines in Fig.~\ref{figmtm}) we
find time evolutions of the imbalance, the phase difference and the velocity which agree very well with
the GPE results (thick lines and dots). For comparison, the original TMA without the correction is shown (dashed
lines).
\newpage


\begin{thebibliography}{99}
\bibitem{foot} In writing this Hamiltonian we are neglecting a term 
$\varepsilon (1-Z^2)\cos(2\Delta\phi)$, which is indeed negligible since $\varepsilon/\Lambda\sim 0.003$.
\end{thebibliography}
\end{document}